\documentclass{article}
\usepackage{fullpage}
\usepackage{amsmath}
\usepackage{amsthm}
\usepackage{amssymb}
\usepackage{url}
\usepackage{graphicx}
\usepackage{verbatim}
\usepackage{url}
\usepackage{graphicx}
\usepackage{tikz}
\usepackage{pgfplots}
\usepackage{fullpage}
\usepackage{amsmath}
\usepackage{amsthm}
\usepackage{amssymb}
\usepackage{url}
\usepackage{graphicx}
\usepackage{verbatim}
\usepackage{multicol}
\usetikzlibrary{decorations.markings}
\usetikzlibrary{calc}
\usepackage{url}
\usepackage{graphicx}
\usepackage{subfigure}
\usepackage{float}
\usepackage{tikz}
\usepackage{pgfplots}
\usetikzlibrary{calc}
\usepgfplotslibrary{fillbetween}
\usetikzlibrary{patterns}
\tikzstyle{mybox} = [draw=black, fill=white,  thick,
    rectangle, inner sep=10pt, inner ysep=20pt]
\tikzstyle{mybox} = [draw=black, fill=white,  thick,
    rectangle, inner sep=2pt, inner ysep=2pt]
\newtheorem{theorem}{Theorem}

\theoremstyle{definition}

\newtheorem{remark}{Remark}
\newtheorem{example}{Example}

\begin{document}

\title{Approximating Bimatrix Nash Equilibrium Via Trilinear Minimax}


\author{Bahman  Kalantari\thanks{Emeritus Professor of Computer Science, Rutgers University ({kalantari@cs.rutgers.edu}).}}

\date{}
\maketitle

\begin{abstract}

The Bimatrix Nash Equilibrium (NE) for $m \times n$ real matrices $R$ and $C$, denoted as the {\it Row} and {\it Column} players, is characterized as follows: Let $\Delta =S_m \times S_n$, where $S_k$ denotes the unit simplex in $\mathbb{R}^k$. For a given point $p=(x,y) \in \Delta$, define $R[p]=x^TRy$ and $C[p]=x^TCy$. Consequently, there exists a subset $\Delta_* \subset \Delta$ such that for any $p_*=(x_*,y_*) \in \Delta_*$, $\max_{p \in \Delta, y=y_*}R[p]=R[p_*]$ and $\max_{p \in \Delta, x=x_* } C[p]=C[p_*]$.

While Nash Equilibrium can be extended to games with multiple players, the computational complexity of bimatrix NE falls within the class of {\it PPAD-complete} problems, a complexity class within NP but not known to be in P (see Papadimitriou \cite{Papa2}, Daskalakis et al. \cite{Papa}, and Chen and Deng \cite{Chen}). Although the von Neumann Minimax Theorem is a special case of bimatrix NE, we introduce a novel extension termed {\it Trilinear Minimax Relaxation} (TMR) with the following implications concerning Nash Equilibrium: Let $\lambda^*=\min_{\alpha \in S_{2}} \max_{p \in \Delta} (\alpha_1 R[p]+ \alpha_2C[p])$ and $\lambda_*=\max_{p \in \Delta} \min_{\alpha \in S_{2}} (\alpha_1 R[p]+ \alpha_2C[p])$.

$\bullet$  $\lambda^* \geq \lambda_*$.

$\bullet$  $\lambda^*$ is computable as a linear programming in $O(mn)$ time, ensuring $\max_{p_* \in \Delta_*}\min \{R[p_*], C[p_*]\} \leq \lambda^*$, meaning that in any Nash Equilibrium it is not possible to have both players' payoffs to exceed $\lambda^*$.

$\bullet$   $\lambda^*=\lambda_*$ if and only if there exists $p^* \in \Delta$ such that $\lambda^*= \min\{R[p^*], C[p^*]\}$. Such a $p^*$ serves as an
approximate Nash Equilibrium. We analyze the cases where such $p^*$ exists and is computable.

$\bullet$  Even when  $\lambda^* > \lambda_*$, we derive approximate Nash Equilibria. Specifically, under conditions that encompass scenarios such as nonnegative payoff matrices, we obtain $\widehat p^* \in \Delta$ satisfying
\begin{center}
$\max_{p_* \in \Delta_*}\min \{R[p_*], C[p_*]\} \leq r^*  \min \{R[\widehat p^*], C[\widehat p^*]\}$,  $r^* \in [1,2)$.
\end{center}

In summary, the aforementioned properties of TMR and its efficient computational aspects underscore its significance and relevance for Nash Equilibrium, irrespective of the computational complexity associated with bimatrix Nash Equilibrium. Finally, we extend TMR to scenarios involving three or more players. Nevertheless, characterizing the cases where $\lambda^*=\lambda_*$ presents a more challenging task.

\end{abstract}



\newpage

\section{Introduction} \label{sec1}

One of the seminal results in Game Theory was established by John von Neumann \cite{Von}, who proved that any bimatrix zero-sum game possesses an equilibrium. Formally, given an $m \times n$ real matrix $A$ and $p = (x, y) \in S_m \times S_n$, where $S_k = \{u \in \mathbb{R}^k : \sum_{i=1}^k u_i = 1, u_i \geq 0 \}$, let $A[p]=A[x, y]=x^TAy$.

\begin{theorem}[\textnormal{von Neumann Minimax Theorem \cite{Von}}]
  There exists $p^*=(x^*, y^*) \in S_{m} \times S_{n}$ such that
  \begin{equation} \label{Neumann}
    \min_{y \in S_{n}} \max_{x \in S_{m}} A[x,y] = \max_{x \in S_{m}} \min_{y \in S_{n}} A[x,y] = A[p^{*}]. \qed
  \end{equation}
\end{theorem}

The game-theoretic interpretation of (\ref{Neumann}) is as follows: Two players, referred to as \textit{Row} and \textit{Column}, simultaneously select indices  $i \in I_1=\{1, \dots, m\}$ and $j \in I_2=\{1, \dots, n\}$, respectively. In this context, Row wins $a_{ij}$ dollars from Column (if $a_{ij}$ is negative, Row loses $|a_{ij}|$ dollars). The sets $I_1$ and $I_2$ represent the sets of \textit{pure actions} or \textit{pure strategies} for the Row and Column players, respectively. Each $(i, j) \in I=I_1 \times I_2$ is considered a \textit{pure strategy profile}. Row and Column players have the option to choose their actions according to a \textit{mixed strategy} with probability vectors $x \in S_m$ and $y \in S_n$, respectively.  According to the von Neumann Minimax Theorem, there exist mixed strategies $p^*=(x^{*}, y^{*}) \in \Delta=S_m \times S_n$ so that the \textit{minimum expected win} for Row is equal to the \textit{maximum expected loss} for Column.

John Nash considered a more general game in which the players, Row and Column, each have their own $m \times n$ real \textit{payoff matrices}, denoted as $R=(r_{ij})$ and $C=(c_{ij})$, respectively. For each $(i,j)$, the matrix entries $r_{ij}$ and $c_{ij}$ represent the \textit{payoff} or \textit{utility} for the respective player when the pure strategy profile $(i,j)$ is selected. When the players choose their actions with a \textit{mixed strategy}, resulting in the composite probability vector $p=(x,y) \in \Delta$, the \textit{expected utility} or \textit{expected payoff} for Row and Column players are denoted as $R[p]$ and $C[p]$, respectively. Nash's profound and brilliant contribution was demonstrating the existence of an equilibrium:

\begin{theorem}[\textnormal{Bimatrix Nash Equilibrium} \cite{Nash}]
  There exists $p_*=(x_*, y_*) \in S_{m} \times S_{n}$ such that
  \begin{equation} \label{NE}
    \max_{x \in S_{m}} R[x, y_*] = R[p_*], \quad \max_{y \in S_{n}} C[x_*,y] = C[p_*]. \qed
  \end{equation}
\end{theorem}

Thus, neither Row nor Column can improve their expected payoff when the other player's probability vector is fixed. In this sense, $p_*$ is a mixed strategy \textit{Nash Equilibrium} (NE), or simply Nash Equilibrium. Nash Equilibria are not necessarily unique.

When the payoff matrices sum to zero, i.e., $R=-C$, von Neumann's minimax is equivalent to the Nash equilibrium. It can be shown von Neumann's Minimax Theorem is closely connected with linear programming duality, see e.g. Dantzig \cite{Dantzig} and  Chv\'atal \cite{Chvatal}. Due to the polynomial time solvability of linear programming, established by Khachiyan \cite{Khachiyan}, it follows that $p^*$ is computable efficiently. On the other hand, computing a Nash Equilibrium $p_*$ remains a challenging problem after decades of research.

Nash Equilibrium, as established by Nash \cite{Nash}, holds more generally for any finite number, $T$, of players, with each player having their own $T$-dimensional payoff matrix. In this article, we initially focus on the bimatrix case of Nash Equilibrium and propose a novel approach to approximate Nash Equilibria by connecting players' expected payoffs through a convex combination. In doing so, we introduce a \textit{Trilinear Minimax Relaxation} (TMR), a theorem of independent interest. Similar to the LP proof of von Neumann's Minimax Theorem, our proof of TMR utilizes linear programming duality to establish a connection between a \textit{minimax} and \textit{maximin} optimization. Unlike the bilinear case, the optimal values of minimax and maximin may be unequal; however, we demonstrate how solving a primal-dual pair of linear programs allows us to compute the minimax value and use it to derive a useful bound for all Nash Equilibria, as well as computing approximations to Nash Equilibria.

We conduct an analysis of the cases of Trilinear Minimax Relaxation (TMR) where the minimax and maximin values coincide and provide insights into the likelihood of these cases. Regardless of whether the minimax and maximin values are equal, we demonstrate the methodology for computing approximate Nash Equilibria and make assertions about their qualities. Additionally, we explore the scaling of the payoff matrices $R$ and $C$ in a manner that allows the computational approach of TMR to yield approximate Nash Equilibria with equal expected payoffs for both players. Such scaling considerations could offer a satisfactory compromise for both players.

A pivotal concept introduced by Papadimitriou \cite{Papa2} is the creation of a novel computational complexity class for Nash Equilibrium computation, termed PPAD (Polynomial Parity Argument in Directed Graphs). PPAD encompasses all search problems that can be reduced to the {\it END OF THE LINE} problem—a search scenario wherein the input is a directed graph with $n$ vertices, representing the amalgamation of cycles and paths totaling $2^n$ vertices. Daskalakis et al. \cite{Papa} detailed this graph, wherein each vertex is denoted by an n-bit string of $0$s and $1$s. The edges of a vertex, denoted by boolean circuits $P$ and $S$ (each with $n$ input and output bits), signify the predecessor and successor vertices of $v$. An {\it unbalanced} vertex lacks either incoming or outgoing edges. Given such a vertex $v$, the END OF THE LINE problem outputs the terminal end of the path. Consequently, a problem is deemed PPAD-complete if it is polynomially reducible to the END OF THE LINE problem.

Notably, PPAD is a subset of NP. Daskalakis et al. \cite{Papa} demonstrated that Nash Equilibrium computation is PPAD-complete for games involving four or more players. Subsequently, they extended this result to 3-player games \cite{Papa}. Chen and Deng \cite{Chen} further established that Nash Equilibrium computation for bimatrix games (2-player games) is PPAD-complete. For additional insights into the complexity of Nash Equilibrium computation, refer to Etessami and Yannakakis \cite{Yanna}. Various algorithms, such as the Govindan-Wilson algorithm \cite{Govind}, have been proposed for computing Nash Equilibria in n-player games. Conversely, Nash Equilibria are known to be Pareto-inefficient. Dubey \cite{Dubey} demonstrated that Nash equilibria in games with smooth payoff functions are generally Pareto-inefficient—a state where no player can alter their strategy to increase their payoffs without detrimentally affecting another player.

Beyond the investigation into the computational complexity of Nash Equilibria, algorithms dedicated to its computation in the bimatrix case have garnered attention. Notably, the Lemke-Howson algorithm, introduced by Lemke \cite{Lemke} and subsequently analyzed by von Stengel \cite{VonS}, demonstrated a worst-case exponential time complexity. A key precondition for the Lemke-Howson algorithm is the non-degeneracy of the game, meaning that the number of pure best responses to a mixed strategy does not surpass the count of pure strategies with a positive probability, as outlined in \cite{VonS}. However, small perturbations can transform a game into a non-degenerate form. The algorithm relies on the notion that a mixed strategy can be labeled by an index corresponding to the player's pure best strategy and subsequent best responses.

The {\it Tableau Method} employed in the Lemke-Howson algorithm offers a means to translate the geometric interpretation of Nash Equilibria as endpoints of a path into a solution for a linear program. In this context, disjoint paths and cycles are construed as the vertices and edges of a polyhedron. Navigating the vertices of the polyhedron involves pivoting. Leveraging Linear Complementarity, the solution for these Nash Equilibria can be derived, as expounded in \cite{VonS}. Despite its ability to find every Nash Equilibrium of the game, the Lemke-Howson algorithm operates with a worst-case exponential time complexity concerning the number of pure strategies. This is due to the pivoting process and traversal of the polyhedron's vertices, as elucidated in \cite{Sava}. Consequently, the Lemke-Howson algorithm is impractical for games where each player may have a large number of actions.

In conclusion, the computation of Nash Equilibria, even in bimatrix games, appears to be inherently intractable. Furthermore, Nash Equilibria are widely recognized as Pareto-inefficient. This realization prompts a meaningful exploration into the development of efficiently computable relaxations that strike a delicate balance. Polynomial-time approximation algorithms emerge as a promising avenue for computing $\varepsilon$-approximate Nash Equilibria, allowing a flexibility of $\varepsilon$ in the maximization problem (\ref{NE}). Noteworthy among these algorithms is the work of Daskalakis et al. \cite{Mehta}, which provides an approximation guarantee of 0.38197. Additionally, the approximation algorithm by Bosse et al. \cite{Bosse} achieves a slightly superior approximation guarantee of 0.36392.

Even considering these $\varepsilon$-approximate Nash Equilibria, as we shall argue, there exists a compelling reason to compute TMR approximations. Such approximations yield higher expected payoffs than any Nash Equilibria, at least for one player, and potentially for both players.

The remaining sections of the article are organized as follows: In Section \ref{sec2}, we introduce a trilinear minimax formulation designed for the approximation of Nash Equilibria in bimatrix games. Subsection \ref{sec2-1} provides a game-theoretic interpretation of the trilinear formulation, while Subsections \ref{sec2-2} and \ref{sec2-3} delve into strategies under pure and random actions, respectively.

Section \ref{sec3}, states and proves the Trilinear Minimax Relaxation (TMR). This theorem serves as a notable generalization of von Neumann's minimax theorem, offering a specific bound on Nash Equilibria and facilitating the derivation of approximations to Nash Equilibria.

Section \ref{sec4} is dedicated to characterizing the solution of TMR and elucidating how it contributes to the generation of approximate Nash Equilibria. Small examples are provided in this section to demonstrate these approximations. Additionally, the section considers the frequency of the types of solutions under an assumption that the payoffs are selected randomly. In Subsection \ref{sec4-1}, it is demonstrated that under a scaling of payoff matrices, it is possible to make TMR optimal solutions result in approximate Nash Equilibria where the payoffs for both players are equal.

Section \ref{sec5} extends TMR to a multilinear minimax relaxation (MMR) tailored for three or more players. This section also explores the possibilities of extending approximate Nash Equilibria for bimatrix games to approximations of Nash Equilibria for three or more players. We end with final remarks.

\section{A Trilinear Minimax Formulation for Bimatrix Games} \label{sec2}

In this section, our objective is to develop a Trilinear Minimax Relaxation (TMR), which can be perceived as a generalization of von Neumann's Minimax Theorem. The subsequent section will employ TMR to compute a relaxation for Nash Equilibria in bimatrix games.

Consider two players, Row and Column, as described earlier, associated with $m \times n$ payoff matrices $R$ and $C$, respectively. Let $\Delta = S_m \times S_n$. For a given $p=(x,y) \in \Delta$ and $\alpha=(\alpha_1, \alpha_2) \in S_2$, define

\begin{equation}
M[\alpha, p]= \alpha_1 R[p]+ \alpha_2 C[p].
\end{equation}

Our goal is to establish a relationship between $\min_{x \in S_2} \max_{p \in \Delta} M[x,p]$ and $\max_{p \in \Delta} \min_{x \in S_2} M[x,p]$. We aim to identify conditions under which equality holds between these two quantities and when it does not. These results will then be employed to compute  approximation of Nash Equilibria. First, we provide a game-theoretic interpretation of TMR.

\subsection{A Game Theoretic Interpretation for Trilinear Minimax} \label{sec2-1}

Consider a game involving three players: Row, Column, and a \textit{Meta-Player}. Players Row and Column are provided with their respective $m \times n$ payoff matrices, denoted as $R$ and $C$. In this game, each of the three players simultaneously makes a choice:

The \textit{Meta-Player} selects one of the players.
Row selects an action $i$ from the set $I_1=\{1, \dots, m\}$.
Column selects an action $j$ from the set $I_2=\{1, \dots, n\}$.

Once these choices are revealed to all participants, depending on which player was selected by the \textit{Meta-Player}, \textit{Meta-Player} loses $r_{ij}$ dollars to Row or $c_{ij}$ dollars to Column.

\subsection{Strategies Under Pure Actions} \label{sec2-2}

\noindent {Strategy for Meta-Player:} If Row is the selected player, the maximum amount Meta-Player loses is $\max \{r_{ij}:(i,j) \in I=I_1 \times I_2\}$. Otherwise, the maximum amount Meta-Player loses is $\max \{c_{ij}:(i,j) \in I\}$. Thus Meta-Player chooses the player that minimizes the worst-case loss.

\noindent {Strategy for Row and Column:}  For each $(i,j)\in  I$, the Meta-Player  will lose at least $\min \{r_{ij}, c_{ij}\}$ dollars.  So the best action pair $(i,j)$ is the one that maximizes Meta-Player's loss, i.e. the players Row and Column compute the best index pair by solving $\max_{(i,j) \in  I} \min \{r_{ij}, c_{ij}\}$.

\subsection{Strategies Under Randomized Actions} \label{sec2-3}

As in the case of zero-sum games, under pure strategies, the two quantities above may differ. Thus, we consider randomized actions. Suppose that the players choose their actions randomly as follows:

\textbf{Meta-Player:} Chooses player Row or Column with probabilities $\alpha_1$ and $\alpha_2$, respectively. Thus, $\alpha = (\alpha_1, \alpha_2) \in S_2$.

\textbf{Row:} Chooses its action with probability vector $x \in S_m$.

\textbf{Column:} Chooses its action with probability vector $y \in S_n$.

Setting $p=(x, y) \in \Delta$, the expected payoffs for Row and Column are, respectively:

\begin{equation}
    R[p] = R[x, y], \quad C[p] = C[x, y].
\end{equation}

\noindent \textbf{Randomized Strategy for Meta-Player:} For each $\alpha=(\alpha_1,\alpha_2) \in S_2$, the worst expected loss is

\begin{equation}
    \max_{p \in \Delta} M[\alpha, p].
\end{equation}

Thus, Meta-Player chooses the players according to the probability vector $\alpha$ that minimizes this loss:

\begin{equation} \label{minimax21}
\min_{\alpha \in S_2} \max_{p \in \Delta} M[x,p].
\end{equation}

\noindent \textbf{Randomized Strategy for Row and Column Players:}
For each $p = (x, y) \in \Delta$, the minimum expected loss for the Meta-Player is the minimum of $\alpha_1 R[p]+\alpha_2 C[p]$ as $\alpha$ ranges over $S_2$. Thus, the best $(i,j) \in I$ is the one that maximizes this:

\begin{equation} \label{maximin21}
\max_{p \in \Delta} \min_{\alpha \in S_2} M[x,p].
\end{equation}

When the optimal values of (\ref{minimax21}) and (\ref{maximin21}) are equal, we shall say $R$ and $C$ are \textit{gap-free}. We will investigate the relationship between (\ref{minimax21}) and (\ref{maximin21}) and when $R$ and $C$ are gap-free.

\section{A Trilinear Minimax Relaxation for Nash Equilibrium} \label{sec3}

We will refer to $M[\alpha, p]= \alpha_1 R[p]+ \alpha_2 C[p]$ as \textit{weighted expected payoff}. Recall $I_1=\{1, \dots, m\}$, $I_2=\{1, \dots, n\}$, and $I=I_1 \times I_2$. Note that

\begin{equation} \label{form1B}
    M[\alpha, p]=  \sum_{(i, j) \in I} \big (\alpha_1 r_{ij} + \alpha_2 c_{ij} \big ) x_i y_j = \alpha_1 R[p] + \alpha_2 C[p] = \alpha_1 \sum_{(i, j) \in I}  r_{ij}x_iy_j +  \alpha_2 \sum_{(i, j) \in I} c_{ij}x_iy_j.
\end{equation}

Let $Q=(q_{ij})$ be an $m \times n$ matrix with $q_{ij} \geq 0$ such that $\sum_{(i,j) \in I} q_{ij}=1$. In particular, the vector of row and column sums of $Q$ are $x=(x_1, \dots, x_m)^T$ and $y=(y_1, \dots, y_n)^T$, respectively, where

\begin{equation} \label{derived}
    x_i=\sum_{j=1}^n q_{ij}, \quad i=1, \dots, m, \quad y_j=\sum_{i=1}^m q_{ij}, \quad j=1, \dots, n.
\end{equation}

Thus, $p=(x,y) \in S_m \times S_n$, referred to as the \textit{derived solution} of $Q$. Set

\begin{equation}
    \widehat R[Q]= \sum_{(i, j) \in I} r_{ij}q_{ij}, \quad
    \widehat C[Q]= \sum_{(i, j) \in I} c_{ij}q_{ij}.
\end{equation}

The above quantities are simply the Hadamard product of $R$ and $C$ with $Q$. The \textit{relaxed weighted expected payoff} is defined as

\begin{equation}
    \widehat M[\alpha, Q]= \alpha_1 \widehat R[Q]+ \alpha_2 \widehat C[Q].
\end{equation}

The following theorem is the main result of the article, demonstrating conditions under which $R$ and $C$ are gap-free and outlining how the relationship between (\ref{minimax21}) and (\ref{maximin21}) can be leveraged to compute approximate solutions to Nash Equilibrium.

\begin{theorem} \label{TMRthm}  {\bf (Trilinear  Minimax Relaxation)}

(I)
\begin{equation} \label{MNEBG}
\lambda^*=\min_{\alpha \in S_2} \max_{p \in \Delta} M[\alpha,p] \geq \lambda_*=\max_{p \in \Delta} \min_{\alpha \in S_2} M[\alpha,p].
\end{equation}

(II) There exists a pair of primal-dual linear programs, referred to as (LP) and (DLP), whose optimal solutions result in  $\alpha^* \in S_2$ and an $m \times n$ matrix $Q^*$ with nonnegative entries summing up to $1$, where

\begin{equation} \label{NMcor1x}
    \lambda^*=\widehat M[\alpha^*, Q^*].
\end{equation}

(III) For each $\alpha \in S_2$ and each $p \in \Delta$,
\begin{equation} \label{NMcor2}
M[\alpha^*, p]  \leq \widehat M[\alpha^*, Q^*] \leq \widehat M[\alpha, Q^*].
\end{equation}

(IV)
If $\Delta_*$ is the set of all  Nash Equilibrium, $p_*$,
\begin{equation} \label{NNNew}
\max_{p_* \in \Delta_*} \min \{R[p_*], C[p_*]\}   \leq  \lambda^*= \min \{ \widehat R[Q^*], \widehat C[Q^*]\}.
\end{equation}

(V)
If $\alpha^* >0$, $\lambda^*=\widehat R[Q^*]= \widehat C[Q^*]$.\\

(VI)
$\lambda^*=\lambda_*$ if and only if there exists $p^* =(x^*, y^*) \in \Delta$ such that $Q^*=x^*{y^*}^T$ is an optimal solution of (DLP). In particular,
$R[p^*]=\widehat R[Q^*]$,  $C[p^*]=\widehat C[Q^*]$ and
\begin{equation} \label{NNNewX}
\max_{p_* \in \Delta_*} \min \{R[p_*], C[p_*]\} \leq  \lambda^*=\min \{R[p^*],  C[p^*]\}.
\end{equation}
\end{theorem}

Before proving the theorem we present Figure \ref{Fig1}, summarizing the significance of the theorem and TMR in deriving an upper bound for the least value of any Nash Equilibrium.

\begin{figure}[htpb]
	\centering
\begin{tikzpicture}[scale=.7]	
\draw (0.0,0.0) -- (7.0,0.0);
\draw (0.0,1.0) -- (7.0,1.0);
\draw (0.0,2.0) -- (7.0,2.0);
\draw (0.0,-1.0) -- (7.0,-1.0);
\draw (0.0,-2.0) -- (7.0,-2.0);
\filldraw (0,2) circle (2pt) node[below] {NE};
\filldraw (7,2) circle (2pt) node[below] {TMR};
\filldraw (5,2) circle (2pt) node[below] {NE};
\filldraw (5,2) circle (2pt) node[below] {NE};
\filldraw (0,1) circle (2pt) node[below] {NE};
\filldraw (7,1) circle (2pt) node[below] {NE};
\filldraw (5,1) circle (2pt) node[below] {TMR};
\filldraw (5,1) circle (2pt) node[below] {TMR};
\filldraw (0,0) circle (2pt) node[below] {NE};
\filldraw (7,0) circle (2pt) node[below] {NE};
\filldraw (3.4,0) circle (2pt) node[below] {TMR};
\filldraw (5,0) circle (2pt) node[below] {TMR};
\filldraw (0,-1) circle (2pt) node[below] {NE};
\filldraw (7,-1) circle (2pt) node[below] {TMR};
\filldraw (2.5,-1) circle (2pt) node[below] {TMR};
\filldraw (5.3,-1) circle (2pt) node[below] {NE};
\filldraw (0,-2) circle (2pt) node[below] {NE};
\filldraw (7,-2) circle (2pt) node[below] {TMR};
\filldraw (3,-2) circle (2pt) node[below] {NE};
\filldraw (5,-2) circle (2pt) node[below] {TMR};
\end{tikzpicture}
	
\caption{The figure depicts various scenarios where expected Nash Equilibria for the two players are compared with the TMR values. The least TMR is denoted as $\lambda^*$. When $\lambda^*=\lambda_*$, one TMR is $R[p^*]$ and the other is $C[p^*]$.} \label{Fig1}
\end{figure}

\begin{proof} (I) and (II): We shall refer to the left-hand-side optimization in (\ref{MNEBG}) as (MINIMAX) and the right-hand-side optimization as (MAXIMIN). Consider (MINIMAX) and the first representation of $M[\alpha, p]$ in (\ref{form1B}). Given a fixed $\alpha \in S_2$, since each term $x_iy_j$ can equal one, by setting $x_i=y_j=1$, it follows that the maximum of $M[\alpha, p]$ over $p \in \Delta$ is the maximum of the terms $(\alpha_1 r_{ij} + \alpha_2 c_{ij})$, $(i,j) \in I$. To find the maximum of these terms, we place a common upper bound, $\delta$, on each such term and then find the minimum of $\delta$ subject to the condition that $\alpha =(\alpha_1, \alpha_2) \in S_2$. Thus, (MINIMAX) is equivalent to the following:

\begin{equation} \label{LP2al}
{\rm (LP)}~~~~~~~\delta^*=\min\{\delta: r_{ij}\alpha_1 + c_{ij}\alpha_2 \leq \delta, \quad \forall (i,j) \in I, \quad
\alpha_1+\alpha_2 =1, \quad \alpha_1, \alpha_2 \geq 0\}.
\end{equation}

Next, consider (MAXIMIN) in (\ref{MNEBG}) and the second representation of $M[\alpha, p]$ in (\ref{form1B}). For a fixed $p=(x,y) \in \Delta$, since $\alpha$ can be $(1,0)$ or $(0,1)$, it follows that the minimum of $M[\alpha, p]$ over $\alpha \in S_2$ is the minimum of $R[p]$ and $C[p]$. To find the minimum of these, we place a common lower bound, $\lambda$, on each of the two summations and then find the maximum of $\lambda$ as $p=(x,y)$ ranges in $\Delta$. This results in a non-linear programming problem with variable terms $x_iy_j$. However, by replacing each term $x_iy_j$ with a single variable $q_{ij}$, and using the constraint that forces $x \in S_{m}$, $y \in S_{n}$, we get
\begin{equation} \label{sumpiB}
\sum_{(i,j) \in I} q_{ij} = \sum_{(i,j) \in I} x_iy_j=\big (\sum_{i=1}^{m}x_i \big )\times \big (\sum_{j=1}^{n} y_j \big)=1.
\end{equation}

Then the (MAXIMIN) and its relaxations, labeled as (DLP), are described as follows:

\begin{multicols}{2}
\begin{minipage}{.6\linewidth}
\begin{align*}
&\lambda_*=\max{\lambda}\\
{\rm (MAXIMIN)}~~&R[p]=\sum_{(i,j) \in I}  r_{ij}x_iy_j  \geq \lambda,\\
&C[p]=\sum_{(i,j) \in I} c_{ij} x_iy_j \geq \lambda,\\
&\sum_{i=1}^{m}x_i=\sum_{j=1}^{n} y_j=1, \quad x_i, y_j\geq 0.\\
\end{align*}
\end{minipage}
\begin{minipage}{.6\linewidth}
\begin{align*}
&\overline \lambda_*=\max{\lambda}\\
{\rm ~~~~~~~~~~~(DLP)}~~&\widehat R[Q]=\sum_{(i,j) \in I}  r_{ij}q_{ij} \geq \lambda,\\
&\widehat C[Q]=\sum_{(i,j) \in I}  c_{ij}q_{ij} \geq \lambda,\\
&\sum_{(i,j)\in I}  q_{ij}=1, \quad q_{ij} \geq 0.\\
\end{align*}
\end{minipage}
\end{multicols}

As (DLP) is a relaxation of (MAXIMIN), it follows that  $ \overline \lambda_* \geq \lambda_*$. Hence to prove (I), it suffices to show (DLP) is the dual of (LP). This will also be used to prove (\ref{NMcor1x}).
Let $A$ be the  constraint matrix corresponding to the inequalities in (LP), excluding nonnegativity constraints.  Thus $A$ has $k=m \times n$ rows and $2$ columns.
Let $e_2 \in \mathbb{R}^2$ and $e_{k} \in \mathbb{R}^{k}$ be vectors of ones. Then (LP) is equivalent to
\begin{equation} \label{LP1B}
\delta^*=\min\{\delta:~A \alpha \leq \delta e_{k},~ e_2^T\alpha=1,~ \alpha \geq 0\}.
\end{equation}
We claim the dual of (\ref{LP1B}) is the following which coincides with (DLP):
\begin{equation} \label{DLP2B}
\max\{\lambda:~A^Tw \geq \lambda e_2,~ e_{k}^Tw =1,~w \geq 0\}.
\end{equation}
To prove the claim we will use the primal-dual  relation for an LP in the standard form having a finite optimal objective (e.g. see [2]).  Introducing the vector of slack variables $s \in \mathbb{R}^{k}$ in (\ref{LP1B}), we get:
\begin{equation} \label{LP2B}
\delta^*=\min\{\delta:~A \alpha-\delta e_{k} +s=0, ~ e_2^T \alpha  =1, ~ \alpha \geq 0, ~s \geq 0\}.
\end{equation}

To turn (\ref{LP2B}) into a standard LP, without loss of generality assume $\delta^* \geq 0$ (otherwise, replacing $\delta$ with $-\delta$, similar arguments apply). Thus, (\ref{LP2B}) can be written as

\begin{equation} \label{LP2Bxx}
\delta^*=\min\{\delta:~A \alpha -\delta e_{k} +s=0, ~ e_2^T \alpha =1, ~ \alpha \geq 0, ~s \geq 0, ~\delta \geq 0\}.
\end{equation}

To write the dual of (\ref{LP2Bxx}), note that in matrix form it is equivalent to the following, where $0_k$ is the zero vector in $\mathbb{R}^k$ and $I_k$ is the $k \times k$ identity matrix:

 \begin{equation}  \label{LP2Byy}
 \min  \bigg \{ (0_{k},1,0_{k})^T
 \begin{bmatrix}
\alpha\\
\delta\\
s
\end{bmatrix}  :
\begin{bmatrix}
A&-e_{n}&I_{k}\\
e_2^T&0&0_{k}^T
\end{bmatrix}
\begin{bmatrix}
\alpha \\
\delta\\
s
\end{bmatrix} =
 \begin{bmatrix}
0_{k}\\
1
\end{bmatrix}, \quad  (\alpha,\delta,s) \geq 0 \bigg \}.
\end{equation}
The dual of (\ref{LP2Byy}), an LP in the standard form, is
\begin{equation}  \label{LPD2Byy}
 \max  \bigg \{ (0_{k},1)^T
 \begin{bmatrix}
w\\
\lambda
\end{bmatrix}  :
\begin{bmatrix}
A^T&e_2\\
-e_{k}^T&0\\
I_{k}&0_k
\end{bmatrix}
\begin{bmatrix}
w\\
\lambda
\end{bmatrix} \leq
 \begin{bmatrix}
0_{k}\\
1\\
0_{k}
\end{bmatrix} \bigg \}.
\end{equation}

This simplifies into the following:
\begin{equation} \label{DLP1B}
\max\{\lambda:~A^Tw+ \lambda e_2 \leq 0, ~ -e_{k}^Tw \leq 1,~w \leq 0\}.
\end{equation}
Replacing $w$ with $-w$, (\ref{DLP1B}) reduces to
\begin{equation} \label{DLP2Bprime}
\max\{\lambda:~A^Tw \geq \lambda e_2,~ e_{k}^Tw  \leq 1,~w \geq 0\}.
\end{equation}
We claim if $(w^*, \lambda^*)$ is an optimal solution of (\ref{DLP2Bprime}),
$e_k^Tw^* =1$. Otherwise, we can scale $w^*$ by a number larger than one, thereby obtaining a better solution for (\ref{DLP2Bprime}). Thus the constraint $e_{k}^Tw  \leq 1$ in (\ref{DLP2Bprime}) can be replaced with $e_{k}^Tw = 1$, proving that the dual of (\ref{LP1B}) is (\ref{DLP2B}). Thus (DLP) is the dual of (LP).

As $\delta^*$ and $\lambda^*$ denote the respective optimal objective values of (LP) and (DLP), we have $\delta^*=\lambda^*$. Thus
\begin{equation}
\lambda^* =\delta^* \geq \overline \lambda_* \geq \lambda_*.
\end{equation}

Substituting $\alpha^*$ for $\alpha$ and $\delta^*$ for $\delta$ in (LP), we get
\begin{equation} \label{eqnnn}
\alpha^*_1r_{ij} + \alpha^*_2c_{ij} \leq \delta^*, \quad \forall (i,j) \in I.
\end{equation}
Multiplying each inequality in (\ref{eqnnn}) by $q^*_{ij}$ and summing them up gives
\begin{equation} \label{eqcor1}
\widehat M[\alpha^*, Q^*] \leq \delta^* \sum_{(i,j) \in I} q_{ij}^*= \delta^*.
\end{equation}
Next substituting $Q^*$  in (DLP) we get
\begin{equation} \label{bbbb}
\widehat R[Q]=\sum_{(i,j) \in I} r_{ij} q^*_{ij} \geq \lambda^*, \quad \widehat C[Q]=\sum_{(i,j) \in I} c_{ij} q^*_{ij} \geq \lambda^*.
\end{equation}
Multiplying the first inequality in (\ref{bbbb}) by $\alpha_1^*$ and the second by
$\alpha_2^*$, then summing them up we get
\begin{equation}
\widehat M[\alpha^*, Q^*] \geq \lambda^*.
\end{equation}
Since $\delta^*=\lambda^*$, (\ref{NMcor1x}) is proved.

(III):  Given $\alpha \in S_2$, multiplying the first inequality in (\ref{bbbb}) by $\alpha_1$, the second inequality by $\alpha_2$, and summing them up gives
\begin{equation} \label{eqcor3}
\widehat M[\alpha ,Q^*] \geq  (\alpha_1+ \alpha_2) \lambda^*= \lambda^*.
\end{equation}
In particular, we get
\begin{equation} \label{eqcor4}
\widehat M[\alpha^*,Q^*] \geq \lambda^*.
\end{equation}
From  (\ref{eqcor4}) and the (\ref{NMcor1x}) the  proof of the right-hand-side inequality in (\ref{NMcor2}) follows. Given any $p =(x,y) \in \Delta$, multiplying each inequality in (\ref{eqnnn}) by $x_iy_j$ and summing them up gives
\begin{equation} \label{eqcor1X}
M[\alpha^*,p] \leq \delta^* \sum_{(i,j) \in I} x_iy_j = \delta^*.
\end{equation}
Since $\delta^*=\lambda^*$, the  proof of the left-hand-side inequality in (\ref{NMcor2}) also follows.

(IV): Given an NE $p_*$, substituting into the left-hand side of (\ref{NMcor2}) gives $M[\alpha^*, p_*] \leq \lambda^*$. Clearly, $M[\alpha^*, p_*] \geq \min\{R[p_*], C[p_*]\}$. Hence, the proof of the left-hand side inequality in (\ref{NNNew}). To prove the right-hand side inequality in (\ref{NNNew}), consider (DLP). We must have $\widehat R[Q^*] \geq \lambda^*$, $\widehat C[Q^*] \geq \lambda^*$. If these inequalities are both strict, we can increase $\lambda^*$, contradicting its optimality. Hence, at least one of the inequalities must be tight.

(V):  If $\alpha^*>0$, complementary slackness condition in LP implies both constraint in (DLP) are tight.

(VI): Consider (MAXIMIN) optimization. By continuity, there exists $p^*$ such that $R[p^*]=\overline \lambda_*$. Since $\lambda^* \geq \overline \lambda_* \geq \lambda_*$, if $\lambda^* = \lambda_*$, it implies $\overline \lambda_*= \lambda_*$. With $Q^*=x^*{y^*}^T$, clearly $R[p^*]=\widehat R[Q^*]$, $C[p^*]=\widehat C[Q^*]$. Finally, (\ref{NNNewX}) follows from (\ref{NNNew}).
\end{proof}

\begin{remark}  As seen in (\ref{LP2al}), (LP) has three variables: $\alpha_1, \alpha_2$, and $\delta$ and $O(mn)$ constraints. In fact, by using the constraint $\alpha_1+\alpha_2=1$, we can eliminate one of the two variables, $\alpha_1$ or $\alpha_2$, transforming it into an LP with two variables. Such an LP can be solved in linear time in the number of its constraints, see Megiddo \cite{Megiddo}. Solving (LP), complementary slackness conditions can be employed to solve (DLP) in the same amount of time. Hence, $\lambda^*$ and $Q^*$ are computable in $O(mn)$ time.
\end{remark}

\section{Characterization of Optimal Solutions of TMR} \label{sec4}

In this section, we characterize optimal solutions of (DLP) and use it to provide approximate Nash Equilibria. By replacing $\lambda$ in (DLP) as the difference of two nonnegative variables and adding slack variables, (DLP) is reducible to an LP in the standard form with three equations, together with nonnegativity constraints. Assuming $R$ and $C$ are not scalar multiples of each other, the resulting (DLP) in standard form will admit a \textit{basic feasible solution}, a solution of a $3 \times 3$ invertible linear system with nonnegative components.

\begin{theorem} \label{appNE}
Assume $(Q^*, \lambda^*)$ is an optimal solution of  {\rm (DLP)}, $\lambda^* \not =0$ and $Q^*$ has the minimal number of nonzero components.  Let $p^*=(x^*,y^*)$ be the corresponding derived solution (see (\ref{derived})).

(1) $Q^*$ has at most two positive entries.

(2) If $Q^*$ has only one positive entry, $p^*$ satisfies $\widehat R[Q^*]=R[p^*]$ and $\widehat C[Q^*]=C[p^*]$.

(3) If $Q^*$ has two positive entries $q^*_{ij}$, $q^*_{i'j'}$, then it must be the case that the following four strict inequalities (or reverse of the four inequalities), referred as {\it dead-end conditions}, are satisfied:
\begin{equation} \label{dead-end}
r_{ij} > r_{i'j'}, \quad r_{ij} > c_{ij}, \quad  c_{i'j'} > r_{i'j'}, \quad c_{i'j'} > c_{ij}.
\end{equation}
Furthermore,  letting $D=(r_{ij}-r_{i'j'})+(c_{i'j'}-c_{ij})$,
\begin{equation} \label{casethree}
q_{ij}^*=\frac{(c_{i'j'}-c_{ij})}{D},  \quad
q_{i'j'}^*= \frac{(r_{ij}-r_{i'j'})}{D}, \quad
\lambda^*= \frac{r_{ij}c_{i'j'}-r_{i'j'}c_{ij}}{D}=\widehat R[Q^*]=\widehat C[Q^*].
\end{equation}

(4)  If $Q^*$ has two positive entries $q^*_{ij}$, $q^*_{i'j'}$ and  $i =i'$ or  $j =j'$,
\begin{equation}
\lambda^*=\lambda_*= \widehat R[Q^*]=R[p^*] = \widehat C[Q^*]=C[p^*].
\end{equation}

(5)  If $Q^*$ has two positive entries $q^*_{ij}$, $q^*_{i'j'}$ and  $i \neq i'$ and  $j \neq j'$, define the {\it square-root solution} as
$\widehat p^* = (\widehat x^*, \widehat y^*) \in \Delta$, where all components of $\widehat x^{*}$, $\widehat y^{*}$ are equal to zero, except for
\begin{equation}
\widehat x^{*}_{i}=\widehat y^*_j= \frac{1}{\rho^*}\sqrt{q_{ij}^*}, \quad
\widehat x^{*}_{i'}=\widehat y^{*}_{j'}= \frac{1}{\rho^*}\sqrt{q_{i'j'}^*}, \quad \rho^*=\sqrt{q_{ij}^*}+ \sqrt{q_{i'j'}^*}.
\end{equation}
Set  $\widehat Q^*=\widehat x^* \widehat {y^*}^T$,
$r^*={\rho^*}^2$,
$\mu^*= \min \{R[p^*]-\widehat R[\widehat Q^*], \quad C[p^*]-\widehat C[\widehat Q^*]\}$.
If $\lambda^* >0$ and $\mu^* \geq 0$,
\begin{equation} \label{eqlams8}
\frac{1}{2} \lambda^* \leq  \frac{1}{r^*} \lambda^* \leq R[\widehat p^*]  \leq \lambda^*, \quad \frac{1}{2} \lambda^* \leq  \frac{1}{r^*} \lambda^* \leq C[\widehat p^*]  \leq \lambda^*, \quad r^* \in [1,2).
\end{equation}

(6)  Let the {\it trivial solution} be
$\overline p^*= (\overline x^*, \overline y^*) \in \Delta$, where
$\overline x^{*}_{i}=\overline y^*_j= \overline x^{*}_{i'}=\overline y^*_{j'}= \frac{1}{2}$ and for  all other indices, the components of $\overline x^{*}$, $\overline y^{*}$ equal to zero. Let $\overline \mu^*= \frac{1}{4} \min \{r_{ij'}+ r_{i'j}, \quad c_{ij'}+ c_{i'j}\}$. If $\lambda^* >0$, $r_{ij}$, $r_{i'j'}$, $c_{ij}$, $c_{i'j'}$ are nonnegative, and $\overline \mu^* \geq 0$,
\begin{equation} \label{eqlams8X}
\frac{1}{4} \lambda^* \leq R[\overline p^*]  \leq \lambda^*, \quad \frac{1}{4} \lambda^* \leq   C[\overline p^*]  \leq \lambda^*.
\end{equation}
\end{theorem}

\begin{proof} (1): Since we assumed $\lambda^* \not =0$, any optimal solution of (DLP) must include $\lambda$. Hence, any optimal solution with the minimum number of nonzero components, $Q^*$, could have at most two positive entries.

(2): If $Q^*$ has only one positive entry, $q_{ij}^*$, it must equal one. Thus, $p^*=(x^*,y^*)$, where $x^*_i=y_j^*=1$ and all other components of $x^*$, $y^*$ are zero. Clearly, $\widehat R[Q^*]=R[p^*]$ and $\widehat C[Q^*]=C[p^*]$.

(3): If the dead-end conditions are not satisfied, either the \textit{row-dominant conditions} hold: $r_{ij} > r_{i'j'}$ and $c_{ij} > c_{i'j'}$ (or the reverse of these inequalities), or the \textit{column-dominant conditions} hold: $r_{ij} > c_{ij}$ and $r_{i'j'} > c_{i'j'}$ (or the reverse of these inequalities). In the row-dominant case, setting $q_{ij}=1$ leads to $\min\{r_{ij}, c_{ij}\} \geq \lambda^*$. This gives an optimal solution with one less positive entry, leading to a contradiction. In the column-dominant case, by choosing the maximum of $\{c_{ij}, c_{i'j'}\}$ and setting $q_{ij}$ or $q_{i'j'}$ equal to one, we again obtain an optimal solution with one less positive entry. Any other case that is not a dead-end is reducible to a row or column-dominant case. Thus, the dead-end conditions hold, and by the complementary slackness condition of linear programming, we get $$r_{ij}q^*_{ij}+r_{i'j'}q^*_{kj}= c_{ij}q^*_{ij}+c_{i'j'}q^*_{kj}= \lambda^*.$$ Using these together with the fact that $q_{ij}^*+q_{i'j'}^*=1$ results in (\ref{casethree}).

(4): If $i=i'$, in the derived solution of $Q^*$, $x_i^*=1$. If $j=j'$, $y_j^*=1$. Thus $\widehat R[Q^*]=R[p^*]$ and $\widehat C[Q^*]=C[p^*]$.

(5): Suppose $i \not = i'$ and $j \not = j'$. Clearly by the definition of $\widehat p^*$, it must lie in  $\Delta$ and
as $\widehat Q^*=\widehat x^* \widehat {y^*}^T$, $R[\widehat p^*] = \widehat R[\widehat Q^*] \leq \lambda^*$ and  $C[\widehat p^*]=\widehat C[\widehat Q^*] \leq \lambda^*$. Since
$\lambda^*= r_{ij}q^*_{ij}+r_{i'j'}q^*_{kj}= c_{ij}q^*_{ij}+c_{i'j'}q^*_{kj}$, from the definition of $\widehat p^*$, $r^*$, $\mu^*$ and assumption that $\mu^* \geq 0$ we get,
$$R[\widehat p^*]= (r_{ij}\widehat x_i^* \widehat y_j^*+r_{i'j'}\widehat x_{i'}^* \widehat y_{j'}^*)+ (r_{ij'}\widehat x_i^* \widehat y_{j'}^*+r_{i'j}\widehat x_{i'}^* \widehat y_j^*)= \frac{1}{r^*} \lambda^* + (r_{ij'}\widehat x_i^* \widehat y_{j'}^*+r_{i'j}\widehat x_{i'}^* \widehat y_j^*) \geq \frac{1}{r^*} \lambda^* + \mu^*.$$
Similarly, $C[\widehat p^*] \geq \frac{1}{r^*} \lambda^* + \mu^*$.  To complete the proof of (\ref{eqlams8}), note that
$$\max \{\sqrt{u_1}+ \sqrt{u_2}:
u_1+u_2= 1, \quad u_1, u_2 \geq 0\}
= \sqrt{2}.
$$
Since $q_{ij}^*$ and $q_{i'j'}^*$ are positive, $r^* \in  [1,2)$.  Hence (\ref{eqlams8}) holds.

(6): The conditions trivially imply $ \lambda^*/4 \leq R[\overline p^*] \leq \lambda^*$ and $ \lambda^*/4 \leq C[\overline p^*] \leq \lambda^*$.
\end{proof}

\begin{remark} \label{remxx}
Once we have computed square-root solution, we may be able to obtain  an {\it improved square-root solution} as follows: Replace $q_{ij}$ with $x_iy_j$ and $q_{i'j'}$ with $x_{i'}y_{j'}$, then replace $x_{i'}$ by $1-x_i$ and $y_{j'}$ by $1-y_j$. This, in turn, requires putting the terms with $q_{ij'}$ and $q_{i'j}$ back. Thus we compute the
largest  $\lambda$ such that the following intersection is nonempty
\begin{align*}
&r_{ij} x_iy_j + r_{ij'}x_i(1-y_j) +  r_{i'j}(1-x_i)y_j+ r_{i'j'} (1-x_i)(1-y_j) \geq \lambda\\
&c_{ij} x_iy_j + c_{ij'}x_i(1-y_j) +  c_{i'j}(1-x_i)y_j+ c_{i'j'} (1-x_i)(1-y_j) \geq \lambda\\
&0 \leq x_i \leq 1, \quad 0 \leq y_j \leq 1.
\end{align*}
This can be achieved using a graphical tool, allowing us to compute a good approximation to the largest value of $\lambda$. The largest value cannot exceed $\lambda^*$. Thus, we start with that value. Theoretically, both inequalities will equal the largest $\lambda$. This is because the derived solution $p^*$ is not gap-free, and hence the optimal solution of the above cannot occur when $x_i$ or $y_j$ are $0$ or $1$ since otherwise the derived solution would be gap-free.
Thus, analytically, we can approximate the largest $\lambda$  using binary search on the values of $\lambda$ and by solving a quadratic equation in each iteration. The improved square-root solution thus gives an approximation where approximate payoffs for both players are the same.  The question of wether or not $\lambda^*=\lambda_*$ appears to be nontrivial because the optimal solution of (MAXIMIN) may occur for a case when more than two $x_i$'s and $y_j$'s are positive.  This also shows the relevance of the computation of $\lambda^*$ as opposed to the computation of $\lambda_*$ since the former is solvable via a linear programming whereas the latter is a nonlinear optimization.
\end{remark}

\begin{remark}

The (DLP) has $O(mn)$ variables. Assuming $\lambda^* \neq 0$, a brute-force approach to solve it is to choose all possible ${mn \choose 2}$ pairs of indices $(i,j)$ and $(i',j')$, considering the corresponding optimal solution restricted to $q_{ij}$ and $q_{i'j'}$, and then picking the best one. Figure \ref{Figdom} indicates the relationship between the four numbers $r_{ij}$, $r_{i'j'}$, $c_{ij}$, and $c_{i'j'}$, providing a sample of four different categories presented as directed graphs. A total of $14$ different cases can occur, and in $12$ of these cases, the optimal solution would result in gap-free cases ($\lambda^*=\lambda_*$).

Assuming that the numbers $r_{ij}$, $r_{i'j'}$, $c_{ij}$, and $c_{i'j'}$ are randomly selected with uniform probability from a set of $N \leq mn$ distinct numbers, we will compute the probability of the dead-end case $r_{ij} > r_{i'j'}$, $r_{ij} > c_{ij}$, and $c_{i'j'} > r_{i'j'}$, $c_{i'j'} > c_{ij}$. This equals the probability of the other dead-end case with inequalities reversed. Suppose the larger of the two numbers $r_{i'j'}$ and $c_{ij}$ is some $k \leq N-1$. There are $2k-1$ ways to select such a pair: $2k-2$ ways to have only one of $r_{i'j'}$ or $c_{ij}$ equal to $k$, plus when both take the value of $k$. Then each of $r_{ij}$ and $c_{i'j'}$ can take any value in $\{k+1, \dots, N\}$. Thus, there are $(2k-1)(N-k)^2$ ways to select the four numbers so that either $r_{i'j'}$ or $c_{ij}$ is $k$, and the other is less than or equal to $k$. Thus, the overall number of ways to get one of the two dead-end cases is
$$2 \sum_{k=1}^{N-1} (2k-1)(N-j)^2 \leq 4 \sum_{k=1}^{N-1} k(N^2 - 2Nk+k^2)=
4 N^2\sum_{k=1}^{N-1} k - 8N  \sum_{k=1}^{N-1} k^2+ 4 \sum_{k=1}^{N-1} k^3 =$$
$$4N^2 \frac{(N-1)N}{2} - 8 N\frac{(N-1)N(2N-1)}{6} + 4\frac{(N-1)^2N^2}{4} \approx (2-\frac{8}{3}+1) N^4 =\frac{1}{3}N^4$$
Dividing this by $N^4$ gives the probability that a random square of four numbers ends up in a dead-end case.  Thus the probability that solving (DLP) results in a gap-free case is  at least $2/3$.

\begin{figure}[htpb]
	\centering
\begin{tikzpicture}[scale=1.5]
\draw[line width=1.pt, postaction={decorate,decoration={markings,mark=at position 0.6 with {\arrow[scale=1]{>}}}}](-8,0) node[above] {$r_{ij}$~~} -- (-7,0) node[above] {$r_{i'j'}$~~};
\draw[line width=1.pt, postaction={decorate,decoration={markings,mark=at position 0.5 with {\arrow[scale=1]{<}}}}] (-7,0) -- (-7,-1);
\draw[line width=1.pt, postaction={decorate,decoration={markings,mark=at position 0.5 with {\arrow[scale=1]{<}}}}](-7,-1)  node[below] {$c_{i'j'}$~~}-- (-8,-1);
\draw[line width=1.pt, postaction={decorate,decoration={markings,mark=at position 0.6 with {\arrow[scale=1]{>}}}}] (-8,-1) node[below]{$c_{ij}$~~} -- (-8,0);

\draw[line width=1.pt, postaction={decorate,decoration={markings,mark=at position 0.6 with {\arrow[scale=1]{>}}}}](-6,0) node[above] {$r_{ij}$~~} -- (-5,0) node[above] {$r_{i'j'}$~~};
\draw[line width=1.pt, postaction={decorate,decoration={markings,mark=at position 0.5 with {\arrow[scale=1]{<}}}}] (-5,0) -- (-5,-1);
\draw[line width=1.pt, postaction={decorate,decoration={markings,mark=at position 0.5 with {\arrow[scale=1]{<}}}}](-5,-1) node[below]{$c_{i'j'}$~~} --   (-6,-1);
\draw[line width=1.pt, postaction={decorate,decoration={markings,mark=at position 0.6 with {\arrow[scale=1]{<}}}}] (-6,-1)  node[below]{$c_{ij}$~~} --  (-6,0);

\draw[line width=1.pt, postaction={decorate,decoration={markings,mark=at position 0.6 with {\arrow[scale=1]{>}}}}]  (-4,0) node[above]{$r_{ij}$~~}  --  (-3,0) node[above]{$r_{i'j'}$~~};
\draw[line width=1.pt, postaction={decorate,decoration={markings,mark=at position 0.5 with {\arrow[scale=1]{<}}}}] (-3,0)   -- (-3,-1);
\draw[line width=1.pt, postaction={decorate,decoration={markings,mark=at position 0.5 with {\arrow[scale=1]{>}}}}](-3,-1) node[below]{$c_{i'j'}$~~} -- (-4,-1);
\draw[line width=1.pt, postaction={decorate,decoration={markings,mark=at position 0.6 with {\arrow[scale=1]{<}}}}] (-4,-1) node[below]{$c_{ij}$~~}--   (-4,0);

\draw[line width=1.pt, postaction={decorate,decoration={markings,mark=at position 0.6 with {\arrow[scale=1]{>}}}}](-2,0) node[above]{$r_{ij}$~~} -- (-1,0) node[above]{$r_{i'j'}$~~};
\draw[line width=1.pt, postaction={decorate,decoration={markings,mark=at position 0.5 with {\arrow[scale=1]{>}}}}] (-1,0) -- (-1,-1);
\draw[line width=1.pt, postaction={decorate,decoration={markings,mark=at position 0.5 with {\arrow[scale=1]{>}}}}](-1,-1) node[below]{$c_{i'j'}$~~} -- (-2,-1);
\draw[line width=1.pt, postaction={decorate,decoration={markings,mark=at position 0.6 with {\arrow[scale=1]{>}}}}] (-2,-1) node[below]{$c_{ij}$~~} -- (-2,0);

\end{tikzpicture}

	\caption{A sample of four categories (left to right): row and column dominant (4 cases), either row or column dominant but not both (8 cases), dead-end (2 cases), and cycle (2 cases). Arrows represent $>$ relation.}
\label{Figdom}
\end{figure}
\end{remark}

\begin{remark}
 The assumption that $\lambda^* \not =0$  is not restrictive nor an important issue. If $\lambda^*=0$, there could be three nonzero $q_{ij}$ in the optimal solution.  In principle, we can remedy this situation by solving (DLP) and if it is found that $\lambda_*=0$, we add small $\varepsilon$ to each entry of $R$ and $C$. There are other ways, e.g. adding a random $\varepsilon$ to each $q_{ij}$ and avoiding solving the (DLP) twice.
\end{remark}

Once we have solved (DLP), computing $Q^*$ and the derived solution $p^*$, it maybe the case that $\lambda^*=\widehat R[Q^*] > R[p^*]$. Even in this case $p^*$ serves as an approximate NE.  To see if $p^*$ can be improved by a better approximation, we compute the trivial  solution as well as the square-root solution and its improved version. We may also compute the best feasible solution, setting each $q_{ij}=1$.   Next we consider a couple of small examples to demonstrate these approximations.

\begin{example} \label{example1} Let $
R=
\begin{bmatrix}
6&5\\
4&8
\end{bmatrix}, \quad
C=
\begin{bmatrix}
2&7\\
6&3
\end{bmatrix}$.
It is easy to check that $p_*=(x_*,y_*)$ with $x_*=(\frac{3}{8}, \frac{5}{8})$, $y_*=(\frac{3}{5}, \frac{2}{5})$
is a Nash Equilibrium.  This gives  $R[p_*]=\frac{28}{5}=5.6$, $C[p_*]=\frac{36}{8}=4.5$.  In TMR, (DLP) is
$$\max\{\lambda:
6 q_{11}+5q_{12}+4q_{21}+8q_{22}  \geq \lambda, \quad
2 q_{11}+7q_{12}+6q_{21}+3q_{22}  \geq \lambda, \quad
q_{11}+q_{12}+q_{21}+q_{22}=1, \quad
q_{ij} \geq 0 \}.$$
By inspection the optimal solution can be checked to be
$q_{12}^*=\frac{5}{7}$, $q_{22}^*=\frac{6}{7}$, $q_{11}^*=q_{21}^*=0$. This gives  $\lambda^*=\frac{41}{7}$. The derived solution is $p^*=(x^*,y^*)$, $x^*=(\frac{5}{7}, \frac{2}{7})$, $y^*=(0,1)$.  We get $R[p^*]=C[p^*]=\lambda^*=5.85$. It is interesting to note that $p^*$ is not a Nash Equilibrium, yet both players receive a better payoff than the Nash Equilibrium $p_*$ above.
\end{example}

\begin{example} \label{example3} Consider the case where $
R=
\begin{bmatrix}
7&1\\
1&2
\end{bmatrix}, \quad
C=
\begin{bmatrix}
3&1\\
1&4
\end{bmatrix}$.
In TMR, (DLP) is
$$\max\{\lambda:
7 q_{11}+q_{12}+q_{21}+2q_{22}  \geq \lambda, \quad
3 q_{11}+q_{12}+q_{21}+4q_{22}  \geq \lambda, \quad
q_{11}+q_{12}+q_{21}+q_{22}=1, \quad
q_{ij} \geq 0 \}.$$
Solving this by inspection, we get
$q_{11}^*=\frac{1}{3}$, $q_{22}^*=\frac{2}{3}$, $q_{12}^*=q_{21}^*=0$,  $\lambda^*=\frac{11}{3}$.

1. The derived solution  is  $p^*=(x^{*}, y^{*})$, $x^*=y^*=(\frac{1}{3}, \frac{2}{3})$. The
corresponding objective value is $\min \{R[p^*], C[p^*]\}=
\min \{19/9, 23/9\}=19/9=2.11$.

2. The trivial solution is $\overline p^*=(\overline x^*, \overline y^*)$, where
$\overline x^*=\overline y^*=(1/2,1/2)$. The corresponding value is
$\min \{11/4,9/4\}=9/4=2.25.$

3. The square-root solution uses $Q^*$. $\widehat p^*=(\widehat x^{*}, \widehat y^{*})$, where $\widehat x^{*}=\widehat y^{*}=(1/(1 +\sqrt{2}),\sqrt{2}/(1+ \sqrt{2})$. This gives  $R[\widehat p^*]=C[\widehat p^*]= (11+2\sqrt{2})/(3+2\sqrt{2})=2.37$.

4.  The improved square-toot solution gives (see Remark \ref{remxx}) gives inequalities $7x_iy_i-x_i-y_i+2 \geq \lambda$ and $5x_iy_i-3x_i-3y_i+4 \geq \lambda$. Using graphical tool we find  that $\lambda=2.3725$ is possible with $x_i \approx.41$, $y_i \approx .41$.

5. In this example the best objective value among  $q_{ij}=1$. Here this value is $3$ with $q_{11}=1$.

Finally, an NE is $p^1_*=(x^1_*,y^1_*)$ with $x^1_*=y^1_*=(1,0)^T$, giving $R[p^1_*]=7, C[p^1_*]=3$. Another NE is   $p^2_*=(x^2_*,y^2_*)$ with $x^2_*=y^2_*=(0,1)^T$, giving $R[p^2_*]=2, C[p^2_*]=4$.  As expected, $\lambda^*$ is an upper bound to the maximum of the minima in the two Nash Equilibria.
\end{example}

\subsection{Scaling of Payoff Matrices} \label{sec4-1}

Consider alternate TMR problems derived via scaling of the payoff matrices $R$ and $C$ by $t$ and $1-t$, respectively, where $t \in (0,1)$. The corresponding expected meta-player's payoff for $\alpha \in S_2$, $p \in \Delta$ becomes

$$M(t)[\alpha,p]=\alpha_1tR[p]+\alpha_2(1-t)C[p].$$

Next, consider the corresponding minimax pair of solutions for $M(t)$, denoted as $(\alpha^*(t),Q^*(t))$. In particular, $(\alpha^*,Q^*)$ in the previous sections corresponds to $t=1/2$. Any Nash Equilibrium $p_*$ remains an equilibrium point for the scaled matrices. However, from Theorem \ref{TMRthm} part (IV) we have,
$$
\min\{t R[p_*], (1-t) R[p_*]\} \leq \min\{t_1 \widehat R[Q^*(t)], (1-t) \widehat C[Q^*(t)]\}.
$$

We may ask: Is there a scaling $t$ for which the corresponding optimal payoffs are equal? An advantage to such a scaling is that it provides a satisfactory compromise for both players.

\begin{theorem} \label{EQP}
Assume $R,C$ have entries in $[0, 1]$, not both identically zero matrix. For a given $t \in [0,1]$, let $Q^*(t)$ be the corresponding optimal solution of TMR. Then there exists $t_* \in (0,1)$ such that,
\begin{equation} \label{Sc}
t_* \widehat R[Q^*({t_*})] = (1-t_*)\widehat C[Q^{*}({t_*})] = \lambda^*({t_*}).
\end{equation}
If $Q^*({t_*})$ is gap-free then $t_*R[p^{*}({t_*})]=(1-t_*)C[p^{*}({t_*})]$, where $p^{*}({t_*})$ is the derived solution of $Q^*({t_*})$.  Otherwise, if $\widehat p^*(t_*)$ is the corresponding square-root solution, both
$t_*R[\widehat p^{*}({t_*})]$ and $(1-t_*)C[\widehat p^{*}({t_*})]$  lie in
$[\rho^* \lambda^*(t), \lambda^*(t)]$, for some $\rho^* \in [0,1]$.
\end{theorem}

\begin{proof}
Proof of the theorem is based on two known results:

(1) If the constraint coefficient for one variable in a linear programming problem changes continuously as a function of a single variable, the optimal objective value changes continuously.

(2) A continuous function from $[0,1]$ into itself has a fixed point.

Scaling $R$ and $C$ by $t$ in $(0,1)$, results in new primal-dual pair. In particular, the scaled (DLP) constraints $\widehat R[Q] \geq \lambda$ and $\widehat C[Q] \geq \lambda$ get replaced with $\widehat R[Q] \geq \lambda/t$ and $\widehat C[Q] \geq \lambda/(1-t)$. Thus in the scaled (DLP)
the variable $\lambda$ changes continuously in $t$.
Now consider the function $f(t) = t \widehat R[Q^*(t)] - (1-t) \widehat C[Q^*(t)]$. By (1), $f(t)$ it is a continuous function of $t$.  On the other hand $f(0) <0$ while $f(1) >0$. By (2), there exists $t_*$ where $f(t_*)=0$.
\end{proof}

For the $2 \times 2$ matrices in Example (\ref{example1}), it is easy to compute such $t_*$ to be $3/10$, giving equal payoffs.

\section{Multilinear Minimax Relaxation for Nash Equilibrium} \label{sec5}
In this section, we consider a multilinear minimax relaxation for an arbitrary number of players. However, for simplicity, we restrict our discussion to the case of three players. The approach can be readily extended to any finite number of players.

For  each $t=1,2,3$, there is a corresponding three-dimensional $n_1 \times n_2 \times n_3$ matrix, $A_t=(a^t_{ijk})$. Let $\Delta=S_{n_1} \times S_{n_2} \times S_{n_3}$.  For $t=1,2,3$, let $I_t=\{1, \dots, n_t\}$. Let $I=I_1 \times I_2 \times I_3$.  Given $p=(x,y,z) \in \Delta$, for $t=1,2,3$ let
$$A_t[p]=\sum_{(i,j,k) \in I}a^t_{ijk}x_iy_jz_k.$$

\begin{theorem} {\rm (Trimatrix Nash Equilibrium)}
There exists $p_*=(x_*, y_*, z_*) \in \Delta$  such that
\begin{equation} \label{NE3}
\max_{p \in \Delta, x=x_*} A_1[p]= A_1[p_*], \quad \max_{p \in \Delta, y=y_*} A_2[p]= A_2[p_*], \quad \max_{p \in \Delta, z=z_*} A_3[p]= A_3[p_*]. \qed
\end{equation}
\end{theorem}
Given $\alpha=(\alpha_1, \alpha_2, \alpha_3) \in S_3$, and $p \in \Delta$, set
$M[\alpha,p]= \alpha_1A_1[p]+\alpha_2A_2[p]+\alpha_3A_3[p]$. Let (MINIMAX) and (MAXIMIN) be respectively defined as:

\begin{equation}
\min_{\alpha \in S_3} \max_{p \in \Delta} M[\alpha,p], \quad \max_{p \in \Delta} \min_{\alpha \in S_3} M[\alpha,p].
\end{equation}

Analogous to the bimatrix case, let (LP) be defined as

\begin{equation} \label{LP2alx}
{\rm (LP)}~~~~~~\delta^*=\min\{\delta: \sum_{t=1}^3a^t_{ijk}\alpha_t, \quad \forall (i,j,k) \in I, \quad
\sum_{t=1}^3 \alpha_t=1, \quad \alpha_t \geq 0\}.
\end{equation}

Given an $n_1 \times n_2 \times n_3$  matrix $Q=(q_{ijk})$, for each $t=1,2,3$, let
$$\widehat A_t[Q]=\sum_{(i,j,k) \in I}a^t_{ijk}q_{ijk}.$$

Analogous to the bimatrix case, the corresponding (MAXIMIN) and its relaxation are defined as

\begin{multicols}{2}
\begin{minipage}{0.5\linewidth}
\begin{align*}
&\lambda^*=\max{\lambda}\\
{\rm (MAXIMIN)}~~~&A_t[p] \geq \lambda, \quad t=1,2,3\\
&\sum_{i=1}^{n_1} x_i=\sum_{j=1}^{n_2} y_j=\sum_{k=1}^{n_3} z_k=1,  \quad x, y,z \geq 0.\\
\end{align*}
\end{minipage}
\begin{minipage}{0.5\linewidth}
\begin{align*}
&\overline \lambda_*=\max{\lambda}\\
{\rm (DLP)}~~~&\widehat A_t[Q] \geq \lambda, \quad t=1,2,3,\\
&\sum_{(i,j,k) \in I}q_{ijk}=1, \quad q_{ijk} \geq 0.\\
\end{align*}
\end{minipage}
\end{multicols}

The following is a generalization of Theorem \ref{TMRthm}, having analogous proof.

\begin{theorem} \label{TMRthmG}  {\bf (Multilinear  Minimax Relaxation)}

(I)
\begin{equation} \label{MNEBGG}
\lambda^*=\min_{\alpha \in S_3} \max_{p \in \Delta} M[\alpha,p] \geq \lambda_*=\max_{p \in \Delta} \min_{\alpha \in S_3} M[\alpha,p].
\end{equation}

(II) By solving a primal-dual of LPs, referred as (LP) and (DLP), from any of their respective optimal solutions  $\alpha^* \in S_3$ and an $n_1\times n_2 \times n_3$ matrix $Q^*$ with nonnegative entries summing up to $1$, we get
\begin{equation} \label{NMcor1xG}
\lambda^*=\widehat M[\alpha^*, Q^*]=\sum_{t=1}^3 \alpha_t^* \widehat A_t[Q^*].
\end{equation}

(III) For each $\alpha \in S_3$ and each $p \in \Delta$ we have
\begin{equation} \label{NMcor2G}
M[\alpha^*, p]  \leq \widehat M[\alpha^*, Q^*] \leq \widehat M[\alpha, Q^*].
\end{equation}

(IV)
If $\Delta_*$ be the set of all  Nash Equilibrium, $p_*$,
\begin{equation} \label{NNNewG}
\max_{p_* \in \Delta_*} \min \{A_t[p_*], t=1,2,3\} \leq \lambda^*=  \min \{ \widehat A_t[Q^*],t=1,2,3\}.
\end{equation}

(V) For $t=1,2,3$, if $\alpha^*_t >0$, $\widehat A_t[Q^*]=\lambda^*$.\\

(VI)
$\lambda^*=\lambda_*$ if and only if there exists $p^* =(x^*, y^*, z^*) \in \Delta$ such that $Q^*=(x_i^*y_j^*z_k^*)$ is an optimal solution of (DLP) and $\widehat A_t[Q^*]
=A_t[p^*]$, $t=1,2,3$. In particular,
\begin{equation} \label{NNNewXX}
\max_{p_* \in \Delta_*} \min \{A_1[p_*], A_2[p_*], A_3[p_*]\} \leq  \min \{A_1[p^*],  A_2[p^*], A_3[p^*]\}. \qed
\end{equation}
\end{theorem}

From (\ref{NNNewXX}), not only one player's expected payoff is $\min \{ A_t[p^*], t=1,2,3\}$, better than one player payoff in any NE, the other players payoff is at least this much.

The following is a generalization of Theorem \ref{appNE} with analogous proof.

\begin{theorem}
Let $(Q^*, \lambda^*)$ be an optimal solution of {\rm (DLP)}, where $Q^*$ has the minimum number of positive entries.  Assume $\lambda^* \not= 0$. Let the derived solution of $Q^*$ be $p^*=(x^*,y^*,z^*) \in \Delta$, where
\begin{equation}
x_i^*= \sum_{j \in I_2, k \in I_3} q^*_{ijk}, \quad y_j^*= \sum_{i \in I_1, k \in I_3} q^*_{ijk}, \quad z_k^*= \sum_{i \in I_1, j \in I_2} q^*_{ijk}.
\end{equation}

(1) There exists at most three distinct triplets $(i,j,k)$, $(i',j',k')$, $(i'',j'',k'')$, where the corresponding entry of $Q^*$ is positive.

(2) If there exists $p^*=(x^*,y^*,z^*) \in \Delta$ such that $Q^*=(x_i^* y_j^*z_k^*)$, then for $t=1,2,3$, $A_t[p^*]=\widehat A_t[Q^*]$.  In particular, this is the case if $Q^*$ has only one positive entry, or two positive entries, say  $q^*_{ijk}$ and $q^*_{i'j'k'}$ with $i=i'$ or $j=j'$ or $k=k'$, or three positive entries where either the first indices are the same, or the second indices are the same, or the third indices are the same.

(3) If $Q^*$ has two positive entries where the first, second and third indices are distinct, as in the bimatrix case we define the square-root solution. Suppose $Q^*$ has three positive entries, $q^*_{ijk}$, $q^*_{i'j'k'}$ and $q^*_{i''j''k''}$,
where all the first indices are distinct, and so are all the second and third indices. Then, we define the {\it cube-root solution}
$\widehat p^* = (\widehat x^*, \widehat y^{*}, \widehat z^{*}) \in \Delta$
as follows.   Let $\rho^*$ be the sum of cube-root of the three positive terms.
Define
\begin{equation}
\widehat x^{*}_{i}= \widehat y^{*}_{j}= \widehat z^{*}_{k}=
\frac{1}{\rho^*}\sqrt[3]{q^*_{ijk}}, \quad \widehat x^{*}_{i'}= \widehat y^{*}_{j'}= \widehat z^{*}_{k'}=
\frac{1}{\rho^*}\sqrt[3]{q^*_{i'j'k'}}, \quad \widehat x^{*}_{i''}= \widehat y^{*}_{j''}= \widehat z^{*}_{k''}=
\frac{1}{\rho^*}\sqrt[3]{q^*_{i''j''k''}}.
\end{equation}
Let
\begin{equation}
r^*= {\rho^*}^3, \quad
\mu^*= \min \{A_t[\widehat p^*]- \widehat A_t[Q^*], \quad t=1,2,3\}.
\end{equation}
If $\lambda^* >0$ and $\mu^* \geq 0$, for $t=1,2,3$ we have,
\begin{equation} \label{eqlams8XX}
\frac{1}{3} \lambda^* \leq  \frac{1}{r^*} \lambda^* \leq A_t[\widehat p^*]  \leq \lambda^*, \quad r^* \in [1,3).
\end{equation}

(4) If $Q^*$ has two positive entries, as in the bimatrix case  we define the {\it trivial solution}. Suppose $Q^*$ has three positive entries $q^*_{ijk}$, $q^*_{i'j'k'}$, $q^*_{i''j''k''}$, where some of the indices are possibly be the same. Let $N_1$ be the number of distinct indices $i,i',i''$ and similarly define $N_2$ and $N_3$ with respect to $j,j',j''$ and $k,k',k''$, respectively.
Let the {\it trivial solution},
$\overline p^* = (\overline x^{*}, \overline y^{*}, \overline z^{*}) \in \Delta$
be defined as follows. Set
\begin{equation}
\overline x^{*}_{i}= \frac{1}{N_1}, \quad \forall i \in I_1, \quad \overline y^{*}_{j}= \frac{1}{N_2}, \quad \forall j \in I_2, \quad  \overline z^{*}_{k}= \frac{1}{N_3}, \quad \forall k\in I_3.
\end{equation}
Let
\begin{equation}
\overline \mu^*= \min \{A_t[\overline p^*]- \widehat A_t[Q^*], \quad t=1,2,3\}.
\end{equation}
If  $\lambda^* \geq 0$ and $\overline \mu^* \geq 0$, for each $t=1,2,3$,
\begin{equation} \label{eqlams8XXX}
\frac{1}{9} \lambda^* \leq \frac{1}{N_1+N_2+N_3} \lambda^* \leq A_t[\widehat p^*]  \leq \lambda^*. \qed
\end{equation}
\end{theorem}

From the above two theorems, we observe that their statements can be extended to any number, $T$, of players. However, as $T$ increases, the complexity of computing $\lambda^*$ also increases. Additionally, the lower bound with respect to the {\it $T$-th root solution} and the {\it trivial solution} changes to $1/T$ and $1/T^2$, respectively. Nevertheless, solving (DLP) remains to be a worthy computation to derive upper bound on the minima of payoffs in any Nash Equilibrium, as well as approximate Nash Equilibria.

\section*{Final Remarks}

In this article, we presented a trilinear minimax formulation for bimatrix games along with its game-theoretic interpretation. We used this to develop a \textit{Trilinear Minimax Relaxation} (TMR), a generalization of von Neumann's Minimax Theorem for zero-sum games. It is noteworthy that, unlike von Neumann's bilinear case, the minimax and maximin values in the trilinear case do not always coincide. However, we showed how to use TMR to derive a relevant bound on Nash Equilibria as well as approximate Nash Equilibria.

Specifically, by solving a primal-dual pair of linear programs, we demonstrated how to calculate an upper bound on the minimum value of the payoffs for both players in any Nash Equilibrium. Furthermore, we showed that when the minimax and maximin values are equal, it is possible to compute expected payoffs, ensuring that one player's payoff is at least as much as the minimum payoff over all Nash Equilibria, and the other player's payoff is at least this much. We provided a characterization of the cases where the minimax and maximin values coincide. Even in situations where these values differ, we outlined schemes for computing approximate Nash Equilibria.

We demonstrated that by scaling the payoff matrices, it is possible to achieve equal expected TMR payoffs for the two players in an optimal solution, providing a compromise that may be satisfactory to both. Additionally, we extended the TMR approach to a more generalized setting, termed the \textit{Multilinear Minimax Relaxation} (MMR), applicable to any finite number of players. MMR serves as a tool for gaining insights into all Nash Equilibria and leads to the computation of approximations to Nash Equilibria. However, the characterization of cases where minimax and maximin values coincide becomes more intricate  even in the case of three players. The introduction of MMR raises new research questions related to Nash Equilibrium, presenting interesting and worthwhile challenges for further investigation.

In summary, TMR serves as a valuable approach for gaining insights into all Nash Equilibria and for computing approximate bimatrix Nash Equilibria. The solution to TMR can serve as a meaningful approximation, surpassing the expected payoffs of Nash Equilibrium for at least one player and possibly even both players. Thus, solving TMR remains relevant even if computation of exact or $\varepsilon$-approximate Nash Equilibrium can be achieved efficiently.

The same principles apply more broadly to the computation of MMR approximations. Notably, even in scenarios with a large number of players, each with a moderate number of strategies, the corresponding (DLP) involves a substantial number of variables and constraints, making its linear program computationally nontrivial.

The goal of this article has been of theoretical nature. Interested practitioners can implement  and test TMR or MMR for their own applications. In the bimatrix case, TMR can be implemented in $O(mn)$ time. However, for any number of players any general-purpose linear programming software can be used to solve TMR or MMR. Clearly, TMR should be much more efficient than computing bimatrix Nash Equilibria using Lemke-Howson algorithm \cite{Lemke}.  The same can be said about using MMR versus such algorithms as Govindan-Wilson algorithm \cite{Govind} for computing Nash Equilibria in multiple players. However,  regardless of the complexity of computing exact or $\varepsilon$-approximate Nash Equilibria, solving TMR or MMR should prove to be relevant and fruitful.

\section*{Acknowledgements} I would like to express my gratitude to Dr. Qiang Liu of The University of Texas at Austin, Computer Science Department, for identifying a flaw in a previous version of this article concerning the relationship between minimax and maximin as defined in this article.

\bigskip

\end{document}